\documentclass[
    ,final            % use final for the camera ready runs
  ]
  {aipproc}

\layoutstyle{6x9}

\newcommand{\kms}{km~s$^{-1}$}

\newcommand{\etal}{{\it et al.}}

\newcommand{\be}{\begin{equation}}
\newcommand{\ee}{\end{equation}}

\newcommand{\arcdeg}{{$^\circ$}}

\newcommand{\mnras}{MNRAS}
\newcommand{\apj}{ApJ}
\newcommand{\aj}{AJ}

\newcommand{\aap}{A\&A}
\newcommand{\apjs}{ApJS}

\begin{document}

\title{The Local Velocity Field}


\author{Karen Masters}{address={Harvard-SAO CfA, 60 Garden Street, MS-20, Cambridge, MA 02138, USA},
,email={kmasters@cfa.harvard.edu}}

\begin{abstract}
We only see a small fraction of the matter in the universe, but the rest gives itself away by the impact of its gravity. The distortions from pure Hubble flow (or peculiar velocities) that this matter creates have the potential to be a powerful cosmological tool, but are also a nuisance for extragalactic astronomers who wish to use redshifts to estimate distances to local galaxies. We provide a quick overview of work on the local peculiar velocity field, discussing both simple spherical infall models, non-parametric modeling using redshifts surveys, and full velocity and density field reconstruction from peculiar velocities. We discuss results from a multiattractor model fit to data from the SFI++ sample of peculiar velocities - the best peculiar velocity data currently available. We also talk about the future of samples for the study of the local velocity field, especially the 2MASS Tully-Fisher (2MTF) survey.
\end{abstract}

\maketitle

%%%%%%%%%%%%%%%%%%%%%%%%%%%%%%%%%%%%%%%%%%%%
%% MAINMATTER
%%%%%%%%%%%%%%%%%%%%%%%%%%%%%%%%%%%%%%%%%%%%

\section{Introduction}

 In an expanding universe there is an eternal battle between expansion trying to pull everything apart and gravity pulling it back together. The distortions from the smooth expansion of the universe that the lumpy distribution of matter creates are known as peculiar velocities. In the linear regime (small overdensities) peculiar velocities, ${\bf v}({\bf x})$, are directly proportional to the underlying gravitational acceleration, ${\bf g}({\bf x})$, through
\be
{\bf v}({\bf x}) = \frac{2}{3 H_0 \Omega_M} f  {\bf g}({\bf x}), \label{eq:GI}
\ee
where $H_0$ is the Hubble constant, $\Omega_M$ is the ratio of the matter density to the critical density, and $f$ is a function describing the rate of growth of structure. It is usually assumed that 
$f \sim \Omega_M^{0.6}$ \citep{P80}, although it has recently been pointed out that $f\sim \Omega_M^{0.55}$ is more accurate in $\Lambda$CDM \citep{L07}.

 Observationally the peculiar velocities of galaxies are measured by taking the difference between the actual recessional velocity and the expected smooth expansion velocity at the distance, $d$, of the galaxy ($v_{\rm pec} = v_{\rm obs} - H_0 d$). Such measurements are difficult and can only be made for galaxies in the nearby universe because distance errors typically range from 10-20\%.

\subsection{Virgo Infall}

 The first models for the velocity field in the local universe considered infall onto the nearest large cluster -- the Virgo cluster. The simplest models for infall make the assumption that the cluster is spherically symmetric. 
Numerical simulations of the growth of structure in a $\Lambda$CDM universe have shown that motion along filaments towards the
elliptical dark matter halos of clusters form the dominant pattern of infall in the universe.
In light of this, the spherical approximation seems like an unjustified simplification, and certainly breaks down close to Virgo. 
However, assuming spherical symmetry has several strong advantages over more complicated distributions of
mass, for example  there is a simple empirical correction into the moderately non-linear regime and the effects of multiple attractors can easily be added.
Under spherical symmetry, Eqn. 1 reduces to
\be
u_{\rm infall} = \frac{1}{3} H_0 r \Omega_0^{0.6} \delta
\ee
(where $\delta$ is the overdensity enclosed at position $r$, and $u_{\rm infall}$ is the peculiar velocity of a galaxy pointing directly towards the attractor). This is extended into the partially non-linear regime using Yahil's ``$\rho ^{1/4}$" approximation \citep{Y85} replacing $\delta$ with $\delta (1 + \delta)^{-1/4}$. The first estimate of the Local Group's infall velocity onto
Virgo using this method, suggested an amplitude of $\sim 250$ \kms ~\citep{P76}. Estimates since then have ranged between a favored value of $\sim 220$ \kms ~\citep{FTS98}
to higher values of $\sim 400$ \kms ~\citep{LSH94}, the significant variation presumably coming from fitting a complicated local velocity field with infall onto only a single attractor.

\subsection{The Great Attractor}

Early in the study of the local peculiar velocity field, it was noted that Virgo could not account for most of the Local Group's motion, which is in a direction $\sim 40^\circ $ from the line
of sight to the cluster and also much too large. The mismatch can be explained if the whole of the Local Supercluster (including the Virgo cluster) is moving towards a direction ($\alpha,\delta) \sim$ (10h, -48\arcdeg) at a speed of $\sim 500$ \nolinebreak \kms. This direction close to the plane of the Milky Way points towards the large mass overdensities of the Hydra-Centaurus supercluster, at a distance of about 45 Mpc, and the (then unknown) Shapley supercluster behind it at $D \sim 200$ Mpc, suggesting immediately that either (or both) of these concentrations of mass may be responsible for a significant fraction of the motion of the Local Group. Early studies of the velocity field in this direction were not much more complex than this simple discussion. \citet{LB87} used $D_n-\sigma$ distances to 400 elliptical galaxies to conclude that the source of motion in this direction could not be the Hydra-Centaurus supercluster since it appeared to partake in the flow and coined the term ``great attractor" (GA) to describe the unknown concentration of mass responsible for the motions. This suggestion of a concentration of mass perhaps hidden from our view by the Milky Way led to a string of attempts to locate it. The Shapley Supercluster was discovered as part of this effort \citep{S89} leading some people to suggest it was the GA, however others favored a closer GA (at roughly the distance of Hydra-Centaurus) and evidence for backside infall at this distance has been put forward (and disputed) by various authors. Observations are ongoing to search for large structures hidden by the Milky Way, and in 2004 an entire conference was devoted to the subject \citep{FL04}. The solution seems likely to lie in a combination of both answers. It is well understood that infall models vastly oversimplify the complicated matter distribution in the real universe, and that the motions of galaxies may well be dominated by infall along filaments. The best fit spherical attractor found in any sample therefore gives only an idea of the center of mass of the distribution traced by the unavoidably biased sample
so the conflict may well result from departures from spherical infall due to the complicated mass distribution in this area, and the likely solution will include infall of some magnitude onto both Hydra-Centaurus and Shapley.

\vspace{0.5cm}

Numerical simulations of the growth of structure in a $\Lambda$CDM universe have shown that motion along filaments towards the elliptical dark matter halos of clusters form the dominant pattern of infall in the universe. However a simple multiattractor model (describing infall onto one or more spherical attractors) can still provide insight into the dominant sources of attraction in the local volume. They also provide a useful parametric correction to distances derived from redshifts nearby. Below we discuss the best currently available multiattractor model fit to data from the SFI++ survey \citep{M05}.

\subsection{Non-parametric modeling}

 Non-parametric modeling of the local velocity field can remove conflicts about the most significant single attractors by allowing for much more complicated mass distributions. They use as an input an all-sky galaxy redshift survey. Under the assumption that galaxies trace mass, this data is used to estimate the inferred mass/gravity field and the galaxy positions and peculiar velocities are iterated until convergence is met. For years, the PSCz density field \citep{B99} was the gold standard in such models, more recently work using the 2MASS Redshift Survey has superseded this \citep{E06}. While such models are non-parametric and clearly more realistic than spherical infall models they do require smoothing on relatively large scales and rest on the assumption that the galaxies selected in the redshift survey trace the mass well. Any redshift survey will have selection effects which will undoubtedly bias this assumption to some extent. PSCz selected IR bright galaxies -- thus missing the most massive ellipticals in the cores of clusters; it also had issues with confusion. 2MRS uses K-band selection so traces stellar mass well, but the survey is rather shallow easily missing massive low surface brightness galaxies. Most recent large redshift surveys like 6dFGS \citep{J04} and SDSS \citep{AM07} do not attempt to cover the whole sky so cannot be used to study the local velocity field in this way.

\subsection{Velocity and Density Field Reconstruction}

Since peculiar velocities (in the linear regime) are directly proportional to the gravitational acceleration and this acceleration is the gradient of the gravitational potential field, in theory the full 3-D velocity and density field can be reconstructed from a peculiar velocity survey which measures just the radial component of the velocities. Techniques which attempt to exploit this property have had varying success in the past, struggling with large errors on the measurement of distances to galaxies, poor statistics and uneven sky coverage. POTENT reconstructions \citep{D99} of the velocity and density fields from different surveys agree broadly, but differ in details and require extremely large smoothing ($\sim$ 12 Mpc) to counter the sparse sampling found even in the largest samples available to date such as Mark III ($N \sim 3000$ \citep{W97}) and SFI ($N \sim 2000$; \citep{G97}), meaning that only the very largest peaks in the density field are recovered. Never-the-less it is this technique which promises to take full advantage of the wealth of information available in peculiar velocities so with improved statistical methods and the next generation of surveys we expect to soon see a re-emergence of its use. A velocity-density reconstruction of SFI++ \citep{S07} is in preparation.

\section{A Multiattractor Model from SFI++}
SFI++ builds and improves on the SFI (spiral field I-band) peculiar velocity catalog published by Giovanelli, Haynes \etal ~in the 1990s \citep{G97}. Data for SFI++ has been collected over the last 15+ years; it includes I-band photometry, HI rotation widths and H$\alpha$ rotation curves, used for the derivation of distances via the Tully-Fisher (TF) relation. The sample has provided TF distances/peculiar velocities for $\sim$5000 spiral galaxies out to $cz \sim$ 10,000 \kms ~ which have been made publicly available \citep{S07} and is now considered the largest reliable survey of peculiar velocities available \citep{FW08}. Figure 1 shows peculiar velocities for SFI++ objects within 15 Mpc of the supergalactic plane.

\begin{figure}
\includegraphics[height=.5\textheight]{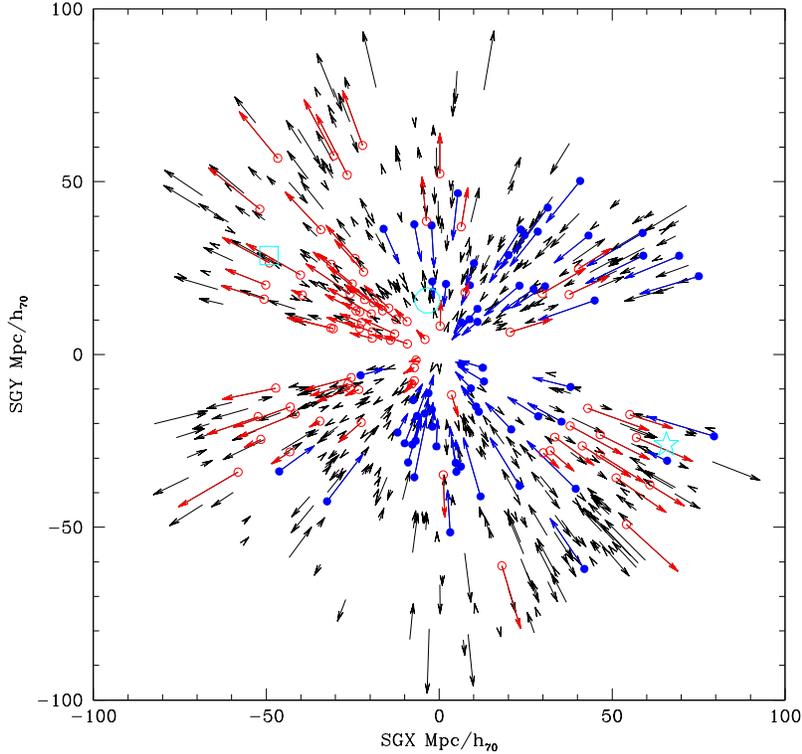}
\caption{SFI++ peculiar velocities are shown for objects within 15 Mpc of the supergalactic plane (SGP); velocities are shown projected onto the SGP. If the velocity has a significant ($> 1\sigma$) positive or negative value it is shown with a open or filled circle respectively. The large open circle indicates the location of the Virgo Cluster, the square shows the center of the Hydra-Centaurus supercluster and the star indicates Pisces-Perseus. \label{masters:sfi++}}
\end{figure}

 The best fit multiattractor model for the SFI++ peculiar velocities \citep{M05} includes infall onto Virgo and the Hydra-Centaurus supercluster, a quadrupolar component describing smaller than average expansion out of the supergalactic plane and a residual dipole pointing towards the Hydra-Centaurus/Shapley superclusters. 
We argue that this model provides evidence for infall onto both the Hydra-Centaurus and Shapley superclusters. Infall onto Shapley, which is significantly outside the sample volume is best fit as a bulk flow and accounts for at least part of the residual dipole. Previous arguments about the nature of the ``great attractor'' can be solved by this physically obvious result. In particular, the mass of the nearby ``great attractor'' associated with Hydra-Centaurus is most likely inflated in models which do not also account for infall onto the Shapley supercluster and is therefore not as significant as suggested in previous ``great attractor'' models \citep{M05}.

This multiattractor model for the local velocity field is fit to a sample almost an order of magnitude larger than any that has been used before \citep{M05}. The SFI++ sample is also being exploited for more complex non-parametric models of the velocity field, where the larger numbers along with high
quality peculiar velocities will provide higher resolution views of the smoothed density field than was possible with previous samples. Meanwhile, this parametric modeling provides the best currently available correction for redshift distances in the local universe.

\section{Future Surveys - 2MTF}

 Previous peculiar velocity surveys have struggled to meet their potential because of large errors on individual measurements, poor statistics and uneven sky coverage. To combat this we are working on the next generation of peculiar velocity surveys. The 2MASS TF survey (2MTF) will provide significantly more uniform sky coverage than has previously been available, making a {\it qualitatively} better sample to study peculiar velocities in the local universe. This survey will make use of existing high quality rotations widths, new HI widths and 2MASS photometry to measure TF distances for all bright inclined spirals in the 2MASS Redshift Survey (2MRS; \cite{H05}). The near-IR selection provides better coverage in the obscured regions near our own Galactic plane which will also aid in studies of the ``great attractor'' region. 

New rotation width observations are required to take advantage of the superior sky coverage provided by selection from 2MRS, particularly in the southern hemisphere and north of the declination range accessible by Arecibo.  In this effort we have been successful obtaining telescope time at both the Green Bank Telescope (GBT) and the Parkes Radio Telescope, and observations are now complete to a fixed K-band magnitude and HI peak flux limit. Data reduction for the almost 1000 galaxies observed at GBT is now significantly complete with a data release paper for this part of the project expected this year. 
For a more in depth description of this project see \citep{M07}.

In the next few years there will be several other new peculiar velocity surveys completed. Peculiar velocity data from the Six Degree Field Galaxy Survey (6dFGS, \citep{J04}) is to be released soon providing peculiar velocities to $\sim$12 000 elliptical galaxies in the southern half of the sky. SFI++ was just released, with data for 5000 spiral galaxies over the whole sky \citep{M06,S07} providing the best currently available peculiar velocity sample (albeit with selection from optical surveys having less than ideal sky coverage). There are also an ever growing number of peculiar velocities from local Type Ia supernova (for example \citep{R04}). The time is ripe for a re-emergence of the use of peculiar velocity data as a cosmological tool. Many of the problems which plagued this field in the 1990s can now be dealt with due to improvements in the catalogs and the statistical tools used to analyze them.

%%%%%%%%%%%%%%%%%%%%%%%%%%%%%%%%%%%%%%%%%%%%%%%%
%% BACKMATTER
%%%%%%%%%%%%%%%%%%%%%%%%%%%%%%%%%%%%%%%%%%%%%%%%

\begin{theacknowledgments}
KLM is a Harvard Postdoctoral Research Fellow supported by NSF grant AST-0406906. I wish to thank all collaborators on the 2MTF project who have helped with GBT and Parkes observing, especially Aidan Crook and Lucas Macri. I also wish to thank numerous collaborators and members (past and present) of the Cornell Extragalactic group without whose work the SFI++ sample would not have been possible.
\end{theacknowledgments}

\end{document}